\documentclass[pre,twocolumn,epsfig,rotate,showpacs]{revtex4}
\usepackage{epsfig}
\usepackage{amsmath}
\usepackage{graphicx}
\usepackage{dcolumn}
\usepackage{bm}

\begin{document}

\title{Normal heat conduction in one-dimensional momentum conserving lattices with asymmetric interactions}
\author{Yi Zhong }
\author{Yong Zhang}
\author{Jiao Wang}
\email{phywangj@xmu.edu.cn }
\author{Hong Zhao}
\email{zhaoh@xmu.edu.cn}
\affiliation{Department of Physics and Institute of Theoretical Physics and Astrophysics,Xiamen University, Xiamen 361005, Fujian, China. }
\date{\today}

\begin{abstract}
We study heat conduction behavior of one-dimensional lattices with asymmetric, momentum conserving interparticle interactions. We find that with a certain degree of interaction asymmetry, the heat conductivity measured in nonequilibrium stationary states converges in the thermodynamical limit. Our analysis suggests that the mass gradient resulting from asymmetric interactions may provide a phonon scattering mechanism in addition to that caused by nonlinear interactions.
\end{abstract}

\maketitle

The heat transport properties of low-dimensional systems have attracted intensive studies for decades \cite{Ried, DL84, DD92, Toda, FPUB, BBB, Prosen2005, Zhao98, Zhao00, Pros, Giar, Giar2005, DaL, Dhar, FPUAB, Nara, Mai, Del, Gray, JSW} (see also Refs. \cite{Dharrev, Lepri, Klages} for reviews and references therein). A challenge is to relate the heat conduction behavior of a system to its microscopic ingredients. In 1984 Casati {\it et al.} investigated the role chaos may play \cite{DL84}, and since then their seminal work has trigged numerous efforts for identifying the microscopic mechanism(s) of the Fourier law. In a one-dimensional (1D) case, the Fourier law states
\begin{equation}
J=-\kappa\frac{\partial T}{\partial x},
\label{Four}
\end{equation}
where $J$ is the heat current, $\frac{\partial T}{\partial x}$ is the spatial temperature gradient, and $\kappa$ is a finite constant termed as ``thermal conductivity." The heat conduction behavior is also known as "normal heat conduction" if it follows the Fourier law or ``abnormal heat conduction" otherwise. Now it has been clarified that chaos is neither sufficient nor necessary to the Fourier law \cite{FPUB, BBB, Prosen2005}.

For 1D lattices, another significant step was made in 1998 by Hu {\it et al.}, who pointed out that, besides the dynamical properties, whether or not the system has a conserved total momentum is another key ingredient \cite{Zhao98,Zhao00}; i.e., lattices with (without) a momentum conservation property should disobey (obey) the Fourier law. In 2000 Prosen and Campbell went a step further; they proved that for 1D momentum conserving lattices with non-vanishing internal pressure the heat conductivity diverges in the thermodynamical limit \cite{Pros}. Though for lattices with a vanishing internal pressure their proof is not applicable, many numerical studies support the same conclusion. In addition, in their later study Prosen and Campbell also showed that momentum conserving is not a necessary condition for abnormal heat conduction \cite{Prosen2005}. More recent progress was made by employing the renormalization group analysis for hydrodynamical models \cite{Nara, Mai} and the mode coupling theory \cite{Lepri, Del, Gray, JSW}. Again both theories predict a divergent heat conductivity in 1D momentum conserving systems. This progress has greatly deepened our understanding of the heat conduction problem. However, in spite of this fact, to our knowledge there are also three counterexamples \cite{Giar, Giar2005, DaL} which are momentum conserving, but instead have convergent heat conductivity. (Reference \cite{DaL}, closely related to the present study, will be discussed in detail later.)

In this paper we investigate the effects of {\it asymmetric} interparticle interactions on the heat conduction of 1D momentum conserving lattices. It has been well known that asymmetric interactions are important for lattice systems; e.g., they may induce a non-vanishing internal pressure and thus the thermal expansion effect \cite{Kittel}. Asymmetric interactions may have significant implications on transport properties as well. This was implied by the studies in Ref. \cite{Pros} and later was well shown in Refs.\cite{Gray, Del}, where two generality classes, corresponding to whether or not the interactions are symmetric, were put forth. In Ref. \cite{Gray}, by proposing a ``mode cascading" relation and incorporating it into the mode coupling theory, the authors theoretically predicted and numerically verified that for systems of symmetric interactions, or equivalently, for systems with equal specific heat capacities at fixed pressure and volume, the bulk viscosity is finite in the thermodynamic limit while for those of asymmetric interactions it is divergent. In addition, the heat conductivity is suggested to diverge in both cases but in different ways. Another development of the mode coupling theory, i.e., the self-consistent mode coupling theory \cite{Del}, has led to a the consistent conclusion. In particular, it concluded that for the 1D momentum conserving systems the heat conductivity would diverge and the divergent exponent is $\frac{1}{3}$ or $\frac{1}{2}$, respectively, for systems with leading cubic or quartic anharmonic nearest neighbor potentials.

In contrast to these results, in the following we shall show that in general momentum conservation does not necessarily imply the breakdown of the Fourier law. Our key finding is the existence of the converged, finite heat conductivity in 1D lattices with asymmetric interparticle interactions. (In the following we refer to ``lattices with asymmetric interparticle interactions" as ``LWAII" for short.) We shall present our simulation results first, then discuss their relation to existing theoretical and numerical studies.

\begin{figure}
\begin{center}
\hskip-0.6cm
\includegraphics[height=4.2cm,width=8.cm]{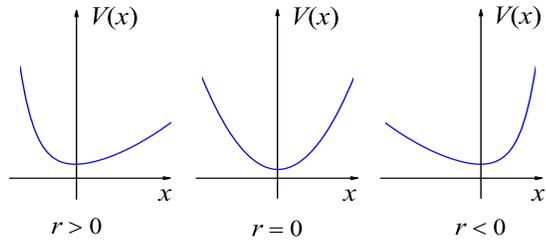}
\vskip-0.4cm
\caption{(Color online) The schematic plot of the potential function $V(x)$ given by Eq. (\ref{Vx}) for $r>0$, $r=0$, and $r<0$, respectively.}
\label{potent}
\end{center}
\end{figure}

We consider homogeneous lattices with nearest neighboring coupling, whose Hamiltonian is
\begin{equation}
H=\sum_{i}\bigr[{\frac{p_{i}^{2}}{2\mu}}+V(x_{i}-x_{i-1}-a)\bigr],
\end{equation}
where $p_{i}$ and $x_{i}$ are,respectively, the momentum and position of the $i$th particle, and $V$ is the potential for the interparticle interaction. As no on-site potentials are involved, this is a momentum conserving model. We assume that the component particles are identical and have a unit mass $\mu$, and the lattice constant, denoted by $a$, is a unit as well. For our aim here the interaction potential with an adjustable asymmetry is favorable. We have investigated several different forms of the interaction potential, some of which will be discussed later, but with all of them qualitatively the same results have been obtained. So as a typical example we shall focus on the following potential:
\begin{equation}
V(x)=\frac{1}{2}(x+r)^{2}+e^{-rx}.
\label{Vx}
\end{equation}
Here $r$ is a controlling parameter that governs the degree of the interaction asymmetry; by increasing $|r|$ from zero where the potential is harmonic and symmetric, one gets increasingly stronger asymmetry. Fixing the system size to be that at zero temperature with a free boundary condition, the potential asymmetry implies a nonzero internal pressure at a finite temperature: While for $r>0$ the internal pressure is positive and the system is thermally expansive, for $r<0$ it is negative and the system is of negative thermal expansion. Note that $x=0$ is the equilibrium point of the potential, and $V(x)$ for $r$ and $-r$ is symmetric with respect to $x=0$. The schematic plots of the potential function are presented in Fig. \ref{potent}.

To measure the heat conductivity of our system, two Nose-Hoover heat baths \cite{NH} at temperatures $T_{L}$ and $T_{R}$ are coupled to the left- and rightmost $N_0$ particles, whose motions follow $\dot x_{i}=\frac{p_{i}}{\mu}$, $\dot p_{i}=-\frac{\partial H}{\partial x_{i}}-\varsigma_{\pm} p_{i}$, and $\dot\varsigma_{\pm} = \frac{p_i^2}{k_B T_{\pm}}-1$. The Boltzmamn constant $k_B$ is set to be unity. The motions of $N$ particles between the heat baths are governed instead by $\dot x_{i}=\frac{p_{i}}{\mu}$ and $\dot p_{i}=-\frac{\partial H}{\partial x_{i}}$.  Given these motion equations, the evolution of the system can be simulated by standard numerical integrating algorithms.

\begin{figure}
\hskip-0.6cm
\includegraphics[angle=0,width=8.0cm,height=5.7cm]{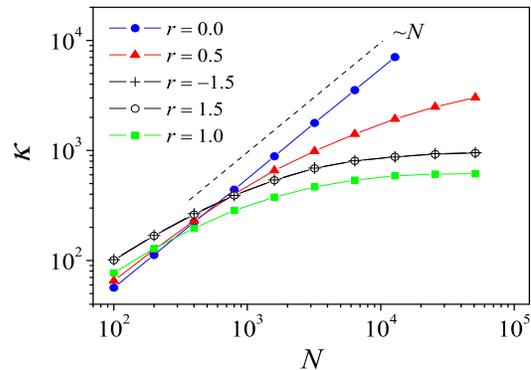}
\vskip-0.4cm
\caption{(Color online) The heat conductivity $\kappa $ vs the number of particles $N$ in our lattice model for various values of the interaction asymmetry parameter $r$. The size and temperatures of the two heat baths coupled to the system are $N_0=12$, $T_{L}=3$, and $T_{R}=2$, respectively. The error bars (not shown) are much smaller than the symbols. The dashed line indicates $\sim N$.}
\label{kappa}
\end{figure}

In our simulations, initially all the particles are assigned to reside on their equilibrium positions with a given random velocity generated from the Maxwellian distribution at an average temperature  $T=\frac{1}{2}(T_{L}+ T_{R})$. Then the system is evolved for a long enough time ($>10^8$ for all the cases investigated) to ensure that it has relaxed to the stationary state. After that the next evolution of time $\sim 10^9$ is performed to obtain the time average of the following quantities: (i) local temperatures $T_{i}\equiv \frac{\langle p_i^2 \rangle}{k_B \mu}$; (ii) local heat currents $J_{i}\equiv \langle\dot{x}_{i} \frac{\partial H}{\partial x_{i}}\rangle$ as adopted conventionally \cite{Zhao98, Mai}; and (iii) heat conductivity $\kappa$ based on
\begin{equation}
\kappa\approx \frac{JNa}{\Delta T}
\label{Eqkp}
\end{equation}
by assuming the Fourier law [see Eq. (\ref{Four})]. Here $J\equiv \langle J_{i}\rangle$ and $\Delta T\equiv T_{L}-T_{R}$. Before we proceed, we emphasize that the numerical results to be presented do not depend on the simulation details considered here. In particular, we have verified that within the error range they do not change if the relaxing time and the average time are increased (by five times), or if we use different definitions of the local heat current, e.g., Eqs. (17) and (23) in Ref. \cite{Lepri} are taken. This is also the case if the leapfrog integrating algorithm mainly adopted in this study is replaced by the Runge-Kutta algorithm of seventh to eighth order.

\begin{figure}
\begin{center}
\hskip-0.6cm
\includegraphics[angle=0,width=8.0cm,height=5.7cm]{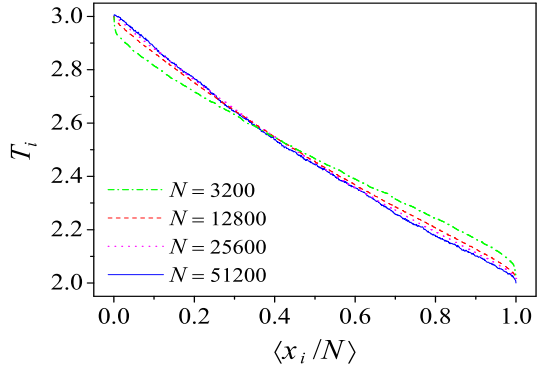}
\vskip-0.4cm
\caption{(Color online) The temperature profiles for $r=1.5$. The size and temperatures of the two heat baths coupled to the system are $N_0=12$, $T_{L}=3$, and $T_{R}=2$, respectively. }
\label{temp}
\end{center}
\end{figure}

Our main results are summarized in Fig. \ref{kappa}, where the dependence of $\kappa$ on the system size $N$ is studied for various values of parameter $r$. The most striking fact revealed there is that for $|r|\ge 1$ the heat conductivity becomes size independent for $N>10^4$, suggesting that the Fourier law holds. This is opposite to the theoretical \cite{Pros, Prosen2005,Nara, Mai, Lepri, Gray, JSW, Del} and simulation \cite{JSW, Dhar, FPUAB} results that in 1D momentum conserving lattices the Fourier law does not hold. To give further support for the converging heat conductivity observed for $|r|\ge 1$, we plot in Fig. \ref{temp} the temperature profiles for $r=1.5$. It shows that for $N>10^4$ the temperature profiles can be well rescaled by $\langle \frac {x_i} {N}\rangle$. This fact justifies the calculation of the thermal conductivity based on Eq. (\ref{Eqkp}) when the system size is sufficient large \cite{Zhao98, Mai, Prosen2005}. In addition, we find that the heat conductivity is the same for $r$ and $-r$, suggesting that in our model thermal expansion and negative thermal expansion have the same implication for heat conduction. As a comparison the heat conductivity for the harmonic chain (with $r=0$) is presented in Fig. \ref{kappa} as well; it diverges linearly with the system size, as expected  \cite{Ried}. We have also studied other asymmetric potentials and found qualitatively the same results. For example, in the case of $V(x)=(1+\lambda)x^2$ for $x\le 0$ and $(1-\lambda) x^2$ for $x>0$, where $0\le|\lambda|<1$ serves as the asymmetry controlling parameter,  $\kappa$ has been observed to saturate in the investigated parameter range $0.5\le\lambda\le 0.8$ when the system size is large enough ($N>10^4$). For this reason we conjecture the finite conductivity is a general existence in 1D LWAII.

Given this one may wonder why previously this was not observed in numerical studies or predicted by theoretical approaches. For the former our analysis suggests the reason could be that the sizes of the systems investigated in previous simulations are not large enough. The transient system size, denoted by $N^*$, where for $N>N^*$ the heat conductivity becomes saturated, is found to depend sensitively on the asymmetry parameters. In the example presented in Fig. \ref{kappa}, we notice that $N^*$ takes its minimal value at $|r|\approx 1$ and increases quickly as $|r|$ becomes smaller. Hence for a less asymmetric potential, e.g., $r=0.5$ (see Fig. \ref{kappa}), a much larger $N^*~(>10^5)$ is expected. Moreover, for a less asymmetric potential, the heat conductivity seems to depend on the system size in a power law for $N\ll N^*$, which explains why in previous studies a power law divergent, rather than a convergent heat conductivity, has been found in 1D LWAII. Indeed, as an example it is easy to check that the asymmetry of the Fermi-Pasta-Ulam(FPU)-$\alpha$-$\beta$ model with $V_{\alpha\beta} (x) = \frac{1}{2}x^{2}+\frac{1}{3} x^{3} + \frac{1}{4} x^{4}$ as was considered in Ref. \cite{FPUAB} is much weaker than the case of $|r|=1$ in our model.

\begin{figure}
\hskip-0.6cm
\includegraphics[angle=0,width=8.1cm,height=5.7cm]{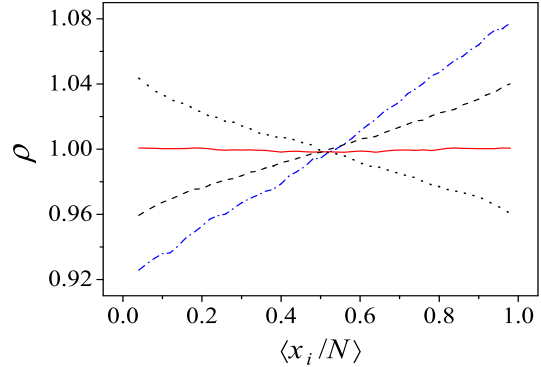}
\vskip-0.4cm
\caption{(Color online) The mass density function, respectively, for our model with $r=1.5$ (dashed line) and $r=-1.5$ (dotted line), FPU-$\beta$ model (solid line), and the variant ding-a-ling model \cite{DaL} (dashed-dotted line). The system size is $N=1200$. Other parameters are $T_{L}=3$, $T_{R}=2$, and $N_0=1$. For the variant ding-a-ling model the natural frequency of the springs connecting the even numbered particles is $\omega_0=1$.}
\label{mass}
\end{figure}

On the other hand, the existing theoretical predictions may not be applicable to the LWAII. It should be noticed that in these theoretical treatments, the system is usually assumed to be at an equilibrium state with a uniform temperature, and thus a homogeneous mass distribution.
But, however, in the LWAII there is an important difference between nonequilibrium stationary states (with a temperature gradient) and equilibrium states: In the former the thermal expansion effect may simultaneously give rise to a mass gradient across the system. This is essentially different from lattices with symmetric interactions where a mass gradient is not expected in either the equilibrium or the nonequilibrium cases. In Fig. \ref{mass} the mass density function $\rho$ for our model is compared with that of the FPU-$\beta$ model with $V_{\beta} (x) = \frac{1}{2}x^{2} + \frac{1}{4} x^{4}$. It shows clearly that, when  being coupled to two heat baths at different temperatures, a mass gradient is eventually established in our system for $r\ne 0$ when the stationary state is approached. It has been known that in systems with symmetric interactions, a nonlinearity of interactions may result in scattering to the heat current that is strong enough to establish the temperature gradient but not strong enough to lead to normal heat conduction \cite{FPUB,Zhao98}. Therefore in systems with asymmetric interactions, the resultant mass gradient may provide an additional scattering mechanism to the heat current. We conjecture that this is the reason why normal heat conduction can then be observed. According to the fluctuation-dissipation theorem, such a macroscopic, nonequilibrium effect must have its microscopic, equilibrium counterpart, but the latter may have not been the object of the existing theoretical studies.

As one more evidence for our conjecture--that the mass gradient may provide an additional scattering mechanism to the heat current--it is worthwhile to notice that in a very recent study \cite{DaL}, a different 1D momentum conserving system having a finite thermal conductivity has also been reported. The system is a momentum conserving variant of the ``ding-a-ling" model \cite{DL84}: The even numbered particles are bound to the adjacent even numbered particles by harmonic springs, and are subject to the elastic collisions with their neighboring odd numbered particles. The odd numbered particles are free except for elastic collisions with their even numbered neighbors.
Significantly, the interactions are {\it asymmetric} due to the elastic collisions, and as a result the system is thermally expansive. In Fig. 4 the mass density of the system in a nonequilibrium state is compared with our system; it can be seen that its asymmetry degree of the interactions is even stronger than our system with $|r|=1.5$. This explains why the saturating regime of the heat conductivity can be numerically accessed in this system \cite{DaL}, provided that our conjecture is correct.

Finally, we would like to emphasize that, not only in 1D LWAII but also in 1D gases, the mass gradient in nonequilibrium stationary state may have significant effects on heat conduction. This has been shown in a 1D hard-core gas with alternative molecule masses \cite{Shunda} and a variant 1D hard-core gas model where two neighboring molecules are bound by a massless string \cite{Gasspressure}. The interparticle interactions in both of them are $asymmetric$. In the former it has been shown both analytically and numerically that, when the system is exposed to two heat baths of different temperatures, the temperature gradient across the system is maintained by the mass density gradient. In the latter the heat conduction behavior has been found to dramatically depend on whether or not the system has a nonzero external pressure (equivalently, a non-zero internal pressure due to the force balance). In addition, as was stressed in Ref. \cite{Gasspressure}, the heat transport properties measured in equilibrium and nonequilibrium states could be qualitatively different. However, it should be noticed that in these two studies the heat conductivity has been shown to diverge in the thermodynamical limit, which implies that, lacking a phonon scattering mechanism in gases, the mass gradient cannot guarantee the Fourier law exclusively.

To summarize, we have performed a numerical investigation for several 1D momentum conserving LWAII and observed normal heat conduction behavior. Comparing our finding in 1D LWAII and the heat conduction characteristics of 1D lattices with symmetric interactions such as the FPU-$\beta$ model, we conjecture that the mass gradient may provide a phonon scattering mechanism in addition to that which is caused by nonlinear interactions, which jointly leads to the observed normal heat conduction behavior in 1D LWAII. Based on our understanding, we conjecture the same mechanism also works in the two-dimensional (2D) case. Indeed, normal heat conduction has been observed in both our ongoing study of 2D momentum conserving LWAII \cite{Zhong} and a recent numerical study of thermal conductivity in empty and water-filled carbon nanotubes \cite{Thomas}. As thermal expansion is ubiquitous among real lattice systems, suggesting their interparticle interactions are generally asymmetric, we expect that the Fourier law generally holds in real low-dimensional systems. In this regard experimental investigations of carbon nanotubes and graphene flakes of large sizes are very desirable.

This work is supported by the NNSF (Grants No. 10805036, No. 10975115, and No. 10925525) and SRFDP (Grant No. 20100121110021) of China.

\end{document}